\documentclass[twocolumn,showpacs,amsmath]{revtex4}

\usepackage[T1]{fontenc}
\usepackage[cp1250]{inputenc}
\usepackage{graphicx}
\usepackage{dcolumn}
\usepackage{bm}
\usepackage{ulem}

\newcommand{\vect}[1]{{\boldsymbol{\mathbf{#1}}}}

\begin{document}

\title{Tailoring quantum superpositions with linearly polarized amplitude-modulated light}

\author{S. Pustelny, M. Koczwara, \L. Cincio, W. Gawlik}
\affiliation{Center for Magneto-Optical Research,
Institute of Physics, Jagiellonian University, Reymonta 4, 30-059
Krak\'ow, Poland}

\date{\today}

\begin{abstract}
Amplitude-modulated nonlinear magneto-optical rotation is a powerful technique that offers a possibility of controllable generation of given quantum states. In this paper, we demonstrate creation and detection of specific ground-state magnetic-sublevel superpositions in $^{87}$Rb. By appropriate tuning of the modulation frequency and magnetic-field induction the efficiency of a given coherence generation is controlled. The processes are analyzed versus different experimental parameters.
\end{abstract}

 \maketitle

\section{Introduction\label{sec:Introduction}}

Quantum coherences are at the heart of many nonlinear and quantum optical phenomena. They are responsible for the appearance of such effects as coherent population trapping \cite{Arimondo1996}, electromagnetically induced transparency \cite{Fleischhauer2005}, extremely slow light propagation \cite{Dutton2004} and its storage \cite{Fleischhauer2000} in a medium, etc. A possibility of controllable generation and modification of an arbitrary quantum state also lays at the very foundations of quantum-state engineering and quantum-information processing (see, for example, Ref.~\cite{Hammerer2009Quantum}).

A specific example of quantum coherence phenomenon is nonlinear magneto-optical rotation (NMOR) \cite{NMORreview,Gawlik2009}. The effect consists of light-intensity-dependent rotation of the polarization plane of linearly polarized light upon its propagation through a medium placed in an external magnetic field. The effect is based on the generation, evolution, and detection of non-equilibrium population distribution and/or quantum coherences between Zeeman sublevels of a given atomic state. In a typical Faraday geometry, in which the magnetic field and light propagation direction are parallel, linearly polarized light generates coherences between Zeeman sublevels differing in the magnetic quantum number $m$ by even values, $\Delta m=2n_i$, where $n_i$ is an integer \cite{FootnoteQuantization}. Despite the fact that different types of coherences can be generated in atoms, it is not usually possible to separate contributions from coherences with particular $\Delta m$ to the NMOR signal that is observed around zero magnetic field, $B\approx 0$ \cite{Lobodzinski1996Multipole,Gateva2007Shape}.

An important breakthrough in the study of NMOR was the application of frequency- \cite{Budker2002FMNMOR} and amplitude-modulated light \cite{Gawlik2006AMOR}, which resulted in the FM NMOR (frequency modulated nonlinear magneto-optical rotation) and AMOR (amplitude modulated optical rotation) techniques. Application of the modulation technique enables the generation of given types of atomic coherences, i.e., coherences between Zeeman sublevels with specific $m$. By exploiting spatial symmetries of the atomic angular-momentum distribution associated with a given quantum state, it is possible to selectively create superpositions between sublevels differing in a magnetic quantum number $m$ by 2, 4, or even more, if only the system supports such coherences \cite{Yashchuk2003Selective,Pustelny2006Pump}. Information about the system's quantum state may also be obtained from the angular momentum distribution by analyzing time-dependent rotation of the polarization plane of light measured at non-zero magnetic field. Thus, in addition to selective generation and detection of the coherences, the technique constitutes a powerful tool in analyzing the evolution of a quantum state of a system. In particular, it allowed detailed investigations of relaxation processes of ground-state coherences in atoms contained in a paraffin-coated vapor cell \cite{Budker2005Antirelaxation,Seltzer2008Testing,Pustelny2006Influence}.

This article presents the investigations on the generation, evolution, and detection of long-living ground-state observables (non-equilibrium population distribution and ground-state coherences, the latter represented by non-diagonal density matrix elements between the ground-state sublevels) in atomic vapor subjected to a longitudinal magnetic field. The measurements are performed in the pump-probe arrangement with one beam used for the creation of atomic polarization and another beam employed for the detection of the system's quantum state. With this arrangement we create the $\Delta m=2$ Zeeman coherences in non-zero magnetic fields. By appropriate tuning of the pumping laser, the coherences are generated in the $F=1$ and $F=2$ hyperfine ground states of $^{87}$Rb. These coherences evolve in the external magnetic field with the frequency determined by the energy splitting between respective sublevels (the Larmor frequency), and are continuously probed with CW light. We study the generation and evolution of the coherences at various pumping and probing conditions. Using magnetic fields such that the nonlinear-Zeeman splitting of the ground-state sublevels is comparable to or exceeds the relaxation rate of this state coherences, we resolve and selectively analyze all three $\Delta m=2$ coherences generated in the $F=2$ state.

The article is organized as follows. The next section presents a theoretical approach demonstrating the relation between ground-state Zeeman coherences and nonlinear magneto-optical rotation with amplitude-modulated light. Section \ref{sec:Experimental} describes experimental apparatus, while Sec.~\ref{sec:Results} presents the results and their analysis. Final remarks are collected in Sec.~\ref{sec:Conclusions}.

\section{Theoretical relation between AMOR signal and ground-state coherences\label{sec:Theory}}

The optical properties of a medium are characterized by the complex refractive index $\eta$
\begin{equation}
    \eta=n+i\kappa=\sqrt{1+\chi},
    \label{eq:ComplexRefractiveIndex}
\end{equation}
where $n$, $\kappa$, $\chi$ denote, respectively, the refractive index, absorption coefficient, and medium electric susceptibility. Knowledge of $\eta$ for different polarizations allows one to determine anisotropic properties of the medium. In particular, the Faraday angle $\varphi$, which is the angle of polarization rotation upon transition through a medium, may be calculated using refractive indices $n_+$ and $n_-$ of two circular polarizations $\sigma^+$ and $\sigma^-$
\begin{equation}
    \varphi=\frac{\omega L}{2c}(n_+-n_-),
    \label{eq:Rotation}
\end{equation}
where $\omega$ is the light frequency, $L$ the length of the medium, and $c$ the speed of light. In order to calculate the susceptibility of a medium, and hence the refractive indices and Faraday angle, the density-matrix formalism may be used \cite{Boyd2003}
\begin{equation}
    \chi_\pm=\frac{N}{E_0}\sum_{m=-F}^{F}d_{mm\pm 1}\rho_{m\pm 1m},
    \label{eq:Chi}
\end{equation}
where $\rho_{m\pm 1m}$ is the optical coherence between the ground-state sublevel $m$ and excited state sublevel $m\pm 1$ and $d_{mm\pm 1}$ is the corresponding dipole matrix element, $N$ the number density of atoms, and $E_0$ is the amplitude of the electric field of the light. Substituting Eq.~(\ref{eq:Chi}) into Eq.~(\ref{eq:ComplexRefractiveIndex}) and expanding the equation into the power series with respect to $\chi$ allows one to link the Faraday angle with the density matrix elements
\begin{widetext}
    \begin{equation}
        \varphi=\frac{\omega L N}{2cE_0}\text{Re}\left(\sum_{m=-F}^{F-1}d_{mm+1}\rho_{m+1m}-\sum_{m=-F+1}^{F}d_{mm-1}\rho_{m-1m}\right).
        \label{eq:RotationFinal}
    \end{equation}
\end{widetext}

In order to calculate the density matrix elements, the evolution of the density matrix $\rho$, governed by the Liouville equation, needs to be considered
\begin{equation}
 \dot{\varrho}=-\frac{i}{\hbar}[H,\varrho]-\frac{1}{2}\{\Gamma,\varrho\}+\Lambda,
 \label{eq:LiouvilleEquation}
\end{equation}
where $H$ is the total Hamiltonian of the system, $\Gamma$ the relaxation operator, $\Lambda$ the repopulation operator describing $\rho$-independent mechanisms such as transit and wall relaxation \cite{FootnoteNeglectingDecay}, and the square and curly brackets denote the commutator and anticommutator \cite{Auzinsh2009Light}. In the considered case, the total Hamiltonian of the system is a sum of the unperturbed Hamiltonian $H_0$ and the Hamiltonians describing interactions of atoms with light $V_l$ and magnetic field $V_B$
\begin{equation}
    H=H_0+V_l+V_B.
    \label{eq:Hamiltonian}
\end{equation}
Assuming the $y$-polarized laser beam, the Hamiltonian describing the light-atom interaction may be written in the dipole approximation as
\begin{equation}
    V_l=-\vect{E}\cdot\vect{d}=-E_0e^{-i\omega t}d_y=-\frac{1}{\sqrt{2}}E_0e^{-i\omega t}(d_-+d_+),
    \label{eq:HamiltonianLight}
\end{equation}
where  $\vect{E}$ is the electric field of the light of amplitude $E_0$, $\vect{d}$ denotes the electric dipole moment operator, and $d_\pm$ are the dipole-matrix elements corresponding to the transitions between the ground- and excited-state Zeeman sublevels differing in the magnetic quantum number $m$ by $\pm 1$.

The magnetic-field interaction Hamiltonian $V_B$ may be presented in the form
\begin{equation}
    V_B=-\vect{\mu}\cdot\vect{B},
    \label{eq:HamiltonianMagnetic}
\end{equation}
where $\vect{\mu}$ is the magnetic dipole moment operator, and $\vect{B}$ the magnetic-field induction. Since in our geometry the magnetic field is applied along the quantization axis, it changes the Zeeman-sublevel energies removing their degeneracy but does not mix them. For an alkali atom, one can calculate the energy shift $\hbar\omega_B$ of a given ground-state magnetic sublevel $m_F$ using the Breit-Rabi formula
\begin{equation}
    \omega_B(m_F)=\frac{E_u}{\hbar}-\frac{\Delta_{HF}}{2(2I+1)}\pm\frac{\Delta_{HF}}{2}\sqrt{1+\frac{4m}{2I+1}x+x^2},
    \label{eq:BreitRabi}
\end{equation}
where $x=\omega_L/\Delta_{HF}$ with $\omega_L=g_F\mu_BB/\hbar$ being the Larmor frequency, $g_F$ the Land\'e factor of the state with a total angular momentum $F$, $\mu_B$ the Bohr magneton, $\Delta_{HF}$ the energy splitting of the ground-state hyperfine doublet, $E_u$ the "center of mass" energy of the level with no hyperfine interaction, and the signs $\pm$ correspond to two hyperfine components $F=I\pm 1/2$.

Combining Eqs.~(\ref{eq:Hamiltonian})-(\ref{eq:BreitRabi}) with Eq.~(\ref{eq:LiouvilleEquation}) allows one to formulate equations describing time evolution of a given density-matrix element $\rho_{a b}$
\begin{equation}
    \dot{\rho}_{a b}=-i\omega_{a b}\rho_{ab}+i\sum_j\left( \Omega_{aj}\rho_{j b}-\rho_{aj}\Omega_{j b}\right)-
    \Gamma_{ab}\left(\rho_{ab}-\rho_{ab}^{eq}\right),
\label{eq:TimeExolutionDMelement}
\end{equation}
where $\Omega_{aj}=E_{aj}d_{aj}/\sqrt{2}\hbar$ is the Rabi frequency associated with the transition between $|a\rangle$ and $|j\rangle$ states, $\omega_{ab}=\omega_B(m_a)-\omega_B(m_b)$ denotes the frequency splitting of the levels, and $\rho_{ab}^{eq}$ the equilibrium value of $\rho_{ab}$.

In order to demonstrate the role of ground-state coherences in rotation of the polarization plane, the explicit formulae for optical coherences need to be written. To derive such analytical formulae, we use the perturbation approach, where the density matrix is expanded into the power series of the electric field of the light
\begin{equation}
    \rho=\sum_{n=0}^\infty\rho^{(n)}E_0^n.
\end{equation}
In such a case, the relation for time evolution of a given density matrix element takes the form
\begin{equation}
    \begin{split}
        \dot{\rho}_{a b}^{(n)}=-&i\omega_{a b}\rho_{ab}^{(n)}+i\sum_j\left( \Omega_{a j}\rho_{j b}^{(n-1)}-\rho_{a j}^{(n-1)}\Omega_{j b}\right)-\\-&\Gamma_{ab}\left(\rho_{ab}^{(n)}-\rho_{ab}^{eq}\right).
    \end{split}
    \label{eq:DMperturbationGeneral}
\end{equation}

Even though in this paper nonlinear magneto-optical rotation with amplitude-modulated light is studied, we first derive analytical formulae for populations, optical and Zeeman coherences when CW light is used (NMOR). Such an approach facilitates the understanding of the problem and provides an insight into the modulated case. By simple generalization of the CW formulae one can write relations for the density-matrix elements when AM light is applied.

\subsection{Low magnetic field, unmodulated light}

Application of the unmodulated light allows one to employ the steady-state approximation  ($\dot{\rho}\equiv 0$). Within this approximation the density-matrix elements calculated in the first three orders of the expansion are given by
\begin{equation}
    \begin{split}
        \rho^{(0)}_{aa}&=\rho^{eq}_{aa},\\
        \sigma^{(1)}_{ab}&=\frac{\Omega_{ab}}{\Delta\omega_{ab}+i\Gamma_{ab}}\rho_{aa}^{(0)},\\
        \rho^{(2)}_{aa}&=-i\frac{\Omega_{ab}\sigma^{(1)}_{ba}-\sigma_{ab}^{(1)}\Omega_{ba}}{\Gamma_{aa}},\\
        \rho^{(2)}_{aa'}&=\frac{\Omega_{ab}\sigma^{(1)}_{ba'}-\sigma_{ab}^{(1)}\Omega_{ba'}}{-\omega_{aa'}+i\Gamma_{aa'}},\\
        \sigma^{(3)}_{ab}&=\frac{\Omega_{ab}\rho_{aa}^{(2)}+\rho_{aa'}^{(2)}\Omega_{a'b}}{\Delta\omega_{ab}+i\Gamma_{ab}},
    \end{split}
    \label{eq:DMperturbation}
\end{equation}
where $\rho_{aa}$ denotes the population of the ground-state sublevel $|a\rangle$, $\sigma_{ab}$ the amplitude of the optical coherence $\rho_{ab}$ ($\rho_{ab}=\sigma_{ab} e^{-i\omega t}$), $\rho_{aa'}$ the ground-state coherence, $\Delta\omega_{ab}=\omega-\omega_{ab}$ the light detuning from the transition between the ground state $|a\rangle$ and the excited state $|b\rangle$, and the superscript to $\rho$ denotes the order of the expansion \cite{FootnoteRWA}. From Eqs.~(\ref{eq:DMperturbation}), the amplitude of the third-order optical-coherence $\sigma_{ab}^{(3)}$ depends on the ground-state Zeeman coherence $\rho_{aa'}^{(2)}$ which is characterized by the ground-state relaxation rate $\Gamma_{aa'}$. Since the ground-state relaxation is much slower than the relaxation of the optical coherence, $\Gamma_{aa'}\ll \Gamma_{ab}$, the third-order optical coherence manifests in the absorption and dispersion via spectral features much narrower than those associated with the first-order coherence. Moreover, Eqs.~(\ref{eq:DMperturbation}) additionally show that not only the widths but also the light-intensity dependences differentiate between the two contributions; while the first-order optical coherence is linear in $\Omega$, and thus is the amplitude of the electric field $E_0$, the third-order coherence depends on $\Omega^3$. Thus, based on the intensity dependences, one demonstrates that the first-order optical coherence is responsible for linear optical phenomena, such as polarization rotation independent of light intensity, and the third-order coherence determines nonlinear phenomena like NMOR and EIT.

As shown above, NMOR is associated with ground-state Zeeman coherences. In particular for the $F=2$ state, three different ground-state coherences ($\rho_{-2,0}$, $\rho_{-1,1}$, $\rho_{0,2}$) contribute to NMOR. Thus, the effect may be used for investigation of the ground-state coherences, in particular, their generation and evolution under interaction with external fields. It is noteworthy, however, that at low magnetic fields independent studies of a given coherence are not possible because of the same dependence of all contributions on the magnetic field and light intensity. Such a distinction would be possible for higher magnetic fields when Zeeman sublevels depend nonlinearly on a magnetic field, however, at such fields the NMOR signals are not observed.

\subsection{Stronger magnetic field, modulated light}

When the intensity of light is sinusoidally modulated, the electric field of the light takes the form
\begin{equation}
    E=\frac{E_0e^{-i\omega t}}{\sqrt{2}}\sqrt{1-a_m\cos\omega_m t},
    \label{eq:IntensityModulation}
\end{equation}
where $a_m$ is the modulation amplitude and $\omega_m$ the modulation frequency. It may be easily shown that for the full modulation ($a_m=1$), relation (\ref{eq:IntensityModulation}) simplifies to $E=E_0\exp(-i\omega t)\sin(\omega_m t/2)$ \cite{FootnoteModulation}. This modification of the light spectrum results in a change of the light-atom interaction Hamiltonian
\begin{equation}
    \begin{split}
        V_l=&-\left(e^{-i(\omega+\omega_m/2)t}+e^{-i(\omega-\omega_m/2)t}\right)\left(d_{-}+d_{+}\right).
    \end{split}
    \label{eq:HamiltonianLightModulated}
\end{equation}

Application of the modulated light rules out the standard steady-state approximation. This is caused by the appearance of the time-dependent Rabi frequency $\Omega=\Omega e^{i\omega_m/2 t}+\Omega e^{-i\omega_m/2 t}$ that drives oscillation of the density-matrix elements at different frequencies. In order to solve Eq.~(\ref{eq:TimeExolutionDMelement}), the density matrix needs to be expanded into the Fourier series of the modulation frequency $\omega_m/2$
\begin{equation}
    \rho=\sum_{k=-\infty}^\infty\rho^{(k)}e^{ik\omega_m/2 t},
    \label{eq:DMExpansion}
\end{equation}
where $\rho^{(k)}$ is the $k$-th Fourier coefficient. Introduction of the Fourier expansion (\ref{eq:DMExpansion}) into Eqs.~(\ref{eq:DMperturbation}) enables application of the steady-state approximation for a given Fourier coefficient of the density matrix. In such a case, one can calculate the time-dependent density matrix elements $\rho^{(l,k)}$ [superscripts $(l,k)$ denote the $l$-th order of the perturbation expansion and the $k$-th order of the Fourier expansion in half the modulation frequency]
\begin{widetext}
    \begin{equation}
        \begin{split}
            \rho^{(0,k)}_{aa}&=\rho^{eq}_{aa}\delta_{k,0},\\
            \sigma^{(1,k)}_{ab}&=\Omega_{ab}\frac{\rho_{aa}^{(0,k-1)}+\rho_{aa}^{(0,k+1)}}{\Delta\omega_{ab}-k\omega_m/2+i\Gamma_{ab}},\\
            \rho^{(2,k)}_{aa}&=\frac{\Omega_{ab}\left(\sigma^{(1,k-1)}_{ba}+\sigma^{(1,k+1)}_{ba}\right)-\Omega_{ba}\left(\sigma_{ab}^{(1,k-1)}+\sigma_{ab}^{(1,k+1)}\right)}
            {-k\omega_m/2+i\Gamma_{aa}},\\
            \rho^{(2,k)}_{aa'}&=\frac{\Omega_{ab}\left(\sigma^{(1,k-1)}_{ba'}+\sigma^{(1,k+1)}_{ba'}\right)-\Omega_{ba'}\left(\sigma_{ab}^{(1,k-1)}-\sigma_{ab}^{(1,k+1)}\right)}
            {-\omega_{aa'}-k\omega_m/2+i\Gamma_{aa'}},\\
            \sigma^{(3,k)}_{ab}&=\frac{\Omega_{ab}\left(\rho_{aa}^{(2,k-1)}+\rho_{aa}^{(2,k+1)}\right)
                +\Omega^{(1)}_{a'b}\left(\rho_{aa'}^{(2,k-1)}+\rho_{aa'}^{(2,k+1)}\right)}{\Delta\omega_{ab}-k\omega_m/2+i\Gamma_{ab}},
        \end{split}
    \label{eq:OpticalCoherenceFourier}
    \end{equation}
\end{widetext}
where $\delta_{lm}$ is the Kronecker delta. Although relations for CW and AM light [Eqs.~(\ref{eq:DMperturbation}) and (\ref{eq:OpticalCoherenceFourier})] are similar there are some significant differences between them. First is the appearance of cross terms that couple different orders of the Fourier expansion. For instance, the $k$-th order populations and Zeeman coherences couple to the $k\pm 1$-th orders of optical coherences. Since the largest density-matrix elements are those with low $k$ (in zeroth order the only non-zero elements are the ground-state populations), the coupling to the higher-order density-matrix elements is weaker. This enables truncation of the formally infinite series (\ref{eq:DMExpansion}) at some finite $k_c$ (usually not bigger than 5). The $k$-$(k\pm 1)$ dependence additionally results in the zeroing of some density matrix elements. It may be shown that populations and Zeeman coherences are nonzero only at even $k$ (for odd $k$ $\rho_{aa'}^{(l,k)}=0$) and the only non-zero optical coherences are these evaluated at odd $k$. The last difference manifests in the appearance of the $-k\omega_m/2$ term in denominators of the formulae for the density-matrix elements, which leads to the generation of additional resonances of the density-matrix elements vs. the modulation frequency. Assuming that the whole energy splitting of the ground-state Zeeman sublevels is due to the magnetic field the resonance arises at non-zero magnetic field; the time-dependent rotation of the polarization plane arises when the Larmor splitting of the levels coincides with a given multiplicity of half of the modulation frequency ($k\omega_m/2=-\omega_{aa'}$). For $F>1$ and low magnetic field all $\Delta m=2$ ground-state coherences are generated with equal efficiency (same energy splitting of the levels) and single AMOR resonance is observed. At stronger fields, i.e., when the nonlinear Zeeman splitting of the sublevels is comparable to, or exceeds the ground-state relaxation rate, each $\Delta m=2$ coherence has different resonance frequency. It is this difference, which allows selective addressing of a given ground-state Zeeman coherence.

In order to calculate dynamic nonlinear magneto-optical rotation, i.e., the AMOR signal, one needs to combine all Fourier coefficients of the density matrix and introduce them into Eq.~(\ref{eq:Chi})
\begin{equation}
    \chi_\pm(t)=\frac{N\text{Tr}\left(\sum_{k=-k_c}^{k_c}\rho^{(k)}e^{ik\omega_m/2t}d_\pm\right)}{E_0\left(e^{-i\omega_m/2 t}+e^{i\omega_m/2 t}\right)}.
    \label{eq:SusceptibilityDymanic}
\end{equation}
Multiplication of Eq.~(\ref{eq:SusceptibilityDymanic}) by $\sin(m\omega_m t)$ or $\cos(m\omega_m t)$ and integrating them over the modulation period gives
\begin{equation}
    \begin{split}
        \chi_{\pm,in}^{(m)}&=\int_0^{2\pi/m\omega_m}\chi_\pm(t)\sin(m\omega_m t),\\
        \chi_{\pm,quad}^{(m)}&=\int_0^{2\pi/m\omega_m}\chi_\pm(t)\cos(m\omega_m t),
    \end{split}
    \label{eq:SusceptibilityInQuad}
\end{equation}
where $m$ is the harmonic number, allows one to find formulae for the in-phase/quadrature amplitude of electric susceptibility at a given harmonic of the modulation frequency. Substituting Eqs.~(\ref{eq:SusceptibilityInQuad}) into Eq.~(\ref{eq:ComplexRefractiveIndex}) and then into Eq.~(\ref{eq:RotationFinal}) enables calculation of the amplitude of time-dependent nonlinear magneto-optical rotation (AMOR signal). It should be noted that the AMOR signal measured in our experiment, i.e. at the first harmonic of the modulation frequency ($m=1$), is described by the third-order optical coherence and $k=1$.

\section{Experimental apparatus\label{sec:Experimental}}

\begin{figure}[h]
    \includegraphics[width=\columnwidth]{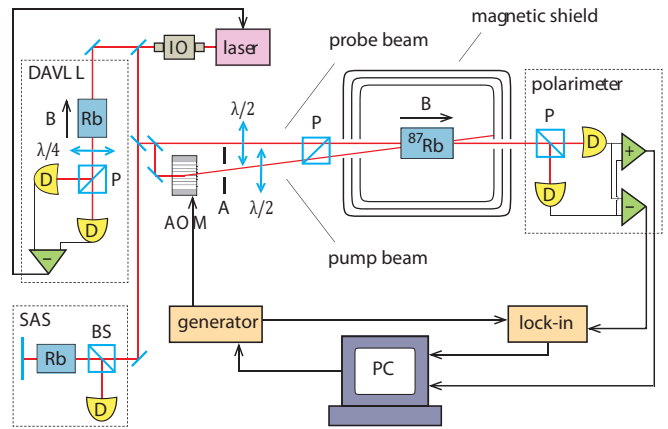}
    \caption{Experimental setup. D is the detector, IO the optical isolator, AOM the acousto-optical modulator, P the polarizer, BS the beam splitter, $\lambda/2$, $\lambda/4$ the half- and quarter-waveplates, respectively, and A is the iris.}
    \label{fig:Setup}
\end{figure}

The layout of the experimental apparatus is shown in Fig.~\ref{fig:Setup}. A paraffin-coated buffer-gas-free cylindrical glass cell of 2-cm in diameter and length of 2 cm contains isotopically enriched sample of $^{87}$Rb. The cell is heated to 50$^\circ$C by a non-magnetic resistive oven providing atomic density of about $5\times 10^{10}$~atoms/cm$^3$ \cite{FootnoteDesorption}. The cell is placed inside a three-layer mu-metal magnetic shield reducing the external, uncontrollable magnetic fields by a factor higher than $10^4$. The residual fields are compensated with two sets of orthogonal magnetic-field coils: one for the first-order magnetic-field gradients and another for the second-order gradients. An additional solenoid is used to generate a highly homogenous and well controlled magnetic field along the light propagation direction which is varied within a range of $\pm 1$~G.

The rubidium atoms interact with two co-propagating, linearly polarized light beams: the pump and the probe. Both beams are generated with the same external-cavity diode laser but their intensities are controlled independently. The laser-light frequency is monitored with a saturated-absorption-spectroscopy system and can be stabilized to a particular transition of the Rb $D1$ line (795~nm) with a dichroic-atomic-vapor laser lock \cite{Corwin1998DAVLL}. The intensity of the pump light is modulated with a single-pass acousto-optical modulator (AOM) optimized for the first order diffraction. Application of AOM enables modulation of light with an arbitrary frequency, amplitude, and waveform. It also leads to a frequency shift of the pump light relative to the probe by 80~MHz. After traversing AOM, the pump light illuminates atoms contained in the vapor cell. The atoms are simultaneously probed with the unmodulated light beam, split off from the main beam before AOM. A balanced polarimeter situated after the shield is employed to analyze the polarization state of the probe. A small angle between the beams allows blocking of the pump before the polarimeter. The polarimeter consists of a Glan polarizer and two photodiodes. The polarimeter differential signal is demodulated with a lock-in amplifier at the first harmonic of the modulation frequency. This signal is than electronically divided by twice the sum of photodiode signals which, for not-too-big rotations, yields information about the amplitude of the polarization rotation [$\varphi\approx(I_1-I_2)/2(I_1+I_2)$, where $I_{1,2}$ are the respective light intensities in the first and second channel of the polarimeter]. Eventually, the signal is stored on a computer which also controls the light modulation and the magnetic-field strength.

\section{Results and discussion\label{sec:Results}}

\begin{figure}[h]
    \includegraphics[width=\columnwidth]{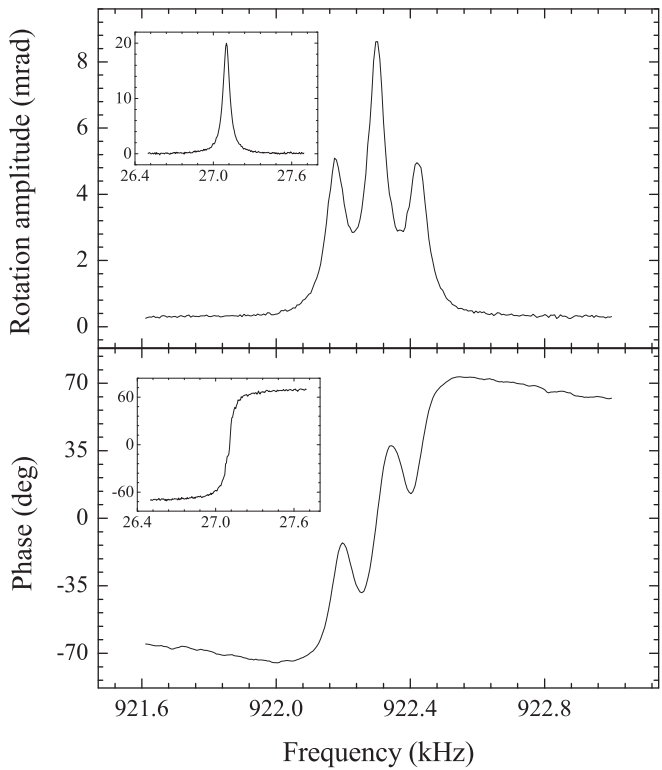}
    \caption{Amplitude and phase of the AMOR signal measured vs. the modulation frequency. For a magnetic field of about 640~mG the AMOR signal is split into three resonances due to the quadratic Zeeman effect. Each resonance corresponds to the different atomic superpositions of the ground-state magnetic sublevels. The insets depict unsplit resonance at $B=19$~mG, where contribution from all superpositions superimpose. Both signals were measured for a pump power of 11~$\mu$W tuned to the center of the $F=2\rightarrow F'=1$ transition and a probe power of 3~$\mu$W.}
    \label{fig:AMORsignal}
\end{figure}

Figure~\ref{fig:AMORsignal} shows the amplitude and phase of the AMOR signal measured vs. the modulation frequency for the pump tuned to the center of the $F=2\rightarrow F'=1$ transition.
For lower magnetic fields, the amplitude dependence is characterized by a single Lorentz resonance and the phase is described by an asymmetric curve, both centered at twice the Larmor frequency (insets to Fig.~\ref{fig:AMORsignal}). For stronger fields, the single AMOR resonance splits into three resonances: the largest central resonance and two smaller resonances shifted symmetrically with respect to the central one. The splitting is also observed at the corresponding phase dependence. Each of the resonances corresponds to different atomic superposition of an individual pair of the ground-state magnetic sublevels.

As discussed in Sec.~\ref{sec:Theory}, the appearance of the AMOR resonance/resonances is associated with generation of the ground-state Zeeman coherences. The process is most efficient when the modulation frequency matches the frequency splitting of the magnetic sublevels differing in the magnetic quantum number $m$ by 2, $\Delta m=2$. In order to calculate the splitting of the sublevels, we expand Eq.~(\ref{eq:BreitRabi}) into the power series of $x$ up to the second order, which for the $F=2$ state of $^{87}$Rb ($I=3/2$) is equal to
\begin{equation}
    \omega_{m,m'}\approx (m-m')\omega_L-(m^2-m'^2)\frac{\omega_L^2}{\Delta_{HF}}.
    \label{eq:MageneticSplitting}
\end{equation}
For three pairs of $\Delta m=2$ sublevels in the $F=2$ state, one obtains
\begin{equation}
    \begin{split}
        \omega_{-2,0}\approx&\ 2\omega_L-4\frac{\omega_L^2}{\Delta_{HF}},\\        \omega_{-1,1}\approx&\ 2\omega_L,\\
        \omega_{0,2}\approx&\ 2\omega_L+4\frac{\omega_L^2}{\Delta_{HF}}.
    \end{split}
    \label{eq:MageneticSplittingThree}
\end{equation}
The first terms in Eqs.~(\ref{eq:MageneticSplittingThree}) arise from the linear Zeeman effect, while the second ones appear due to the quadratic Zeeman effect. For weak magnetic fields, the contribution from the nonlinear effect is significantly smaller than the ground-state relaxation rate ($4\omega_L^2/\Delta_{HF}\ll \gamma$). In that case, the AMOR resonances associated with individual coherences overlap and appear as a single resonance (insets to Fig.~\ref{fig:AMORsignal}). For stronger magnetic fields, the sublevel-splitting frequencies differ sufficiently and the separation of the resonances is observed, as seen in Fig.~\ref{fig:AMORsignal}. The amplitude of the given resonance is determined by the amplitude of the corresponding coherence and appropriate dipole matrix elements. Thus using the NMOR signal and also the relations~(\ref{eq:OpticalCoherenceFourier})-(\ref{eq:SusceptibilityInQuad}) determination of the amplitude of the coherence is possible.

\begin{figure}[htb!]
 \includegraphics[width=\columnwidth]{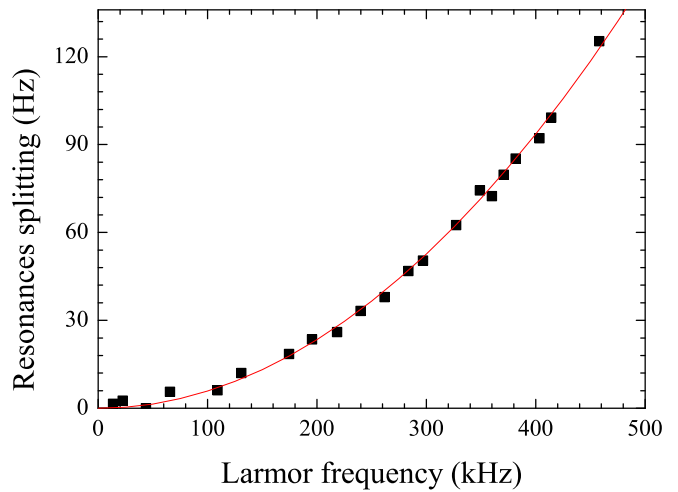}
 \caption{(Color online) Average splitting of the AMOR resonances, $(\omega_r-\omega_l)/2$, measured vs. the Larmor frequency (square).  The observed dependences is in very good agreement with theoretical curve plotted based on Eq.~(\ref{eq:MageneticSplittingThree}) (solid line) and the data from Ref.~\cite{Steck87Rb}. The data were measured with a single sinusoidally modulated light beam of 8-$\mu$W power acting as pump and probe simultaneously.}
\label{fig:NonlinearZeeman}
\end{figure}

In Fig.~\ref{fig:NonlinearZeeman}, the measured splitting of the resonances is presented as a function of the magnetic field.
For low magnetic fields, the resonances are unresolved and no splitting is measured \cite{FootnoteMeasuringSplitting}. The splitting becomes measurable for fields corresponding to Larmor frequencies above a few tens of kHz and increases quadratically with $B$ and $\omega_L$, which is in good agreement with predictions of Eqs.~(\ref{eq:MageneticSplittingThree}).

\begin{figure}[htb!]
    \includegraphics[width=\columnwidth]{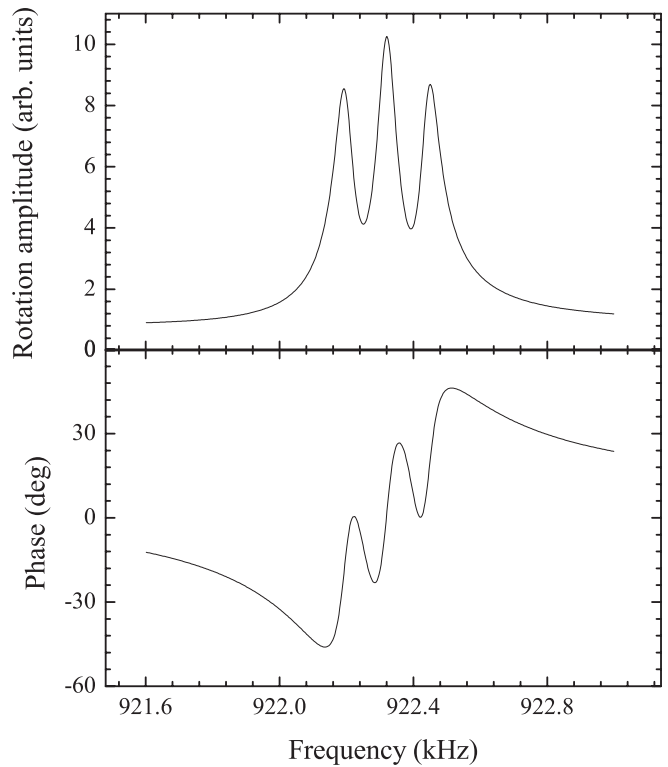}
    \caption{Amplitude and phase of the AMOR signal simulated for $F=2\rightarrow F'=1$ transition using Eqs.~(\ref{eq:OpticalCoherenceFourier}). The signal reveals all the salient features of real signal (Fig.~\ref{fig:AMORsignal}), i.e. resonance splitting, similar amplitude relations.}
    \label{fig:AMORsimulated}
\end{figure}

In order to verify the model developed in Sec.~\ref{sec:Theory}, we simulate AMOR signals for the first harmonic of the modulation frequency, $\omega_m\approx 920$~kHz and a magnetic field of 640~mG (Fig.~\ref{fig:AMORsimulated}).
The simulations reveal all the features observed experimentally (Fig.~\ref{fig:AMORsignal}). For a magnetic field inducing significant nonlinear Zeeman splitting of the levels, three AMOR resonances associated with $\rho_{-2,0}$, $\rho_{-1,1}$, and $\rho_{0,2}$ are observed. The strongest resonance is related to the coherence between the $|-1\rangle$ and $|1\rangle$ sublevels. The remaining two resonances are equally split with respect to the central one and have equal amplitudes. Also the phase of the simulated signal follows the experimentally measured dependence crossing zero at the center of the largest signal. We attribute the observed deviation of the simulated from the measured dependences due to higher order processes, such as power broadening and saturation, which are not included in our model.

\begin{figure}[htb!]
 \includegraphics[width=\columnwidth]{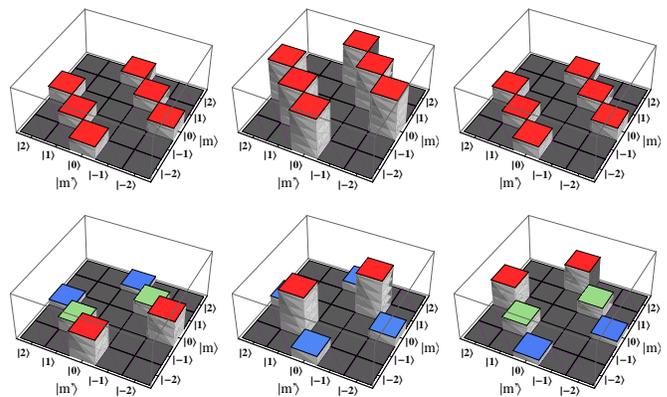}
 \caption{(Color online) Calculated absolute values of the density-matrix elements ($|\rho_{m,m'}|$) of the $F=2$ ground under interaction with linearly-polarized, AM light tuned to the $F=2\rightarrow F'=1$ transition. The top row corresponds to a magnetic field of about $10$~mG, when energy splittings of all $\Delta m=2$ sublevels are equal and the bottom row to a much stronger field of 640~mG causing the AMOR-resonance splittings of 120~Hz, that is four times the ground-state relaxation rate. The central column shows the density-matrix elements for $\omega_m=2\omega_L$, while the side columns correspond to $\omega_m=2\omega_L\pm 3\gamma$, respectively.}
\label{fig:DMsimulations}
\end{figure}

The splitting of the AMOR resonance into three resonances illustrates a possibility of selective addressing of a particular $\Delta m=2$ ground-state coherence. Figure~\ref{fig:DMsimulations} presents simulations of the density matrix at the first harmonic of the modulation frequency calculated at lower and higher magnetic fields corresponding to single and split resonances (Fig.~\ref{fig:AMORsimulated}). The top row is calculated for 10~mG, whereas the bottom for 640~mG, that corresponds to resonance splitting of 120~Hz, $(\omega_r-\omega_l)/2\approx 4\gamma$. The simulations are performed for the modulation frequencies equal to $2\omega_L-3\gamma$ (left column), $2\omega_L$ (middle column), and $2\omega_L+3\gamma$ (right column). As shown in the top row, exact tuning of the modulation frequency to twice the Larmor frequency results in generation of all $\Delta m=2$ coherences with the highest and equal efficiency, while detuning it away uniformly reduces the amplitudes of all of them. In stronger fields the situation is different (bottom row). While for $\omega_m=2\omega_L$ the $\rho_{-1,1}$ coherence is generated with the highest efficiency, the two other coherences are created significantly less efficient. One can selectively increase the amplitude of either of these coherences by appropriate tuning of the modulation frequency. For $\omega_m=2\omega_L-3\gamma$ (left column), the $\rho_{-2,0}$ coherence is most efficiently generated, while for $\omega_m=2\omega_L+3\gamma$ (right column), the $\rho_{0,2}$ coherence has the strongest amplitude. This dependence of the amplitude of the specific coherence on the modulation frequency proves the possibility of selective addressing and the control of specific $\Delta m=2$ coherences.

The AMOR technique is a powerful tool in the analysis of a quantum state of the system. In particular, scanning the modulation frequency, fitting the data with three Lorentz curves, and taking into account strengths of the specific transitions, allows one to extract information about amplitudes of the density-matrix elements. This measurement, however, requires modulation-frequency scan within a range strongly exceeding the splitting of the resonances. In order not to scan the modulation frequency, one may perform free-induction decay measurements, where information about the coherence amplitude is extracted from time-dependent rotation signal (for more details see Ref.~\cite{Acosta2008Production}).

\begin{figure}[htb!]
    \includegraphics[width=\columnwidth]{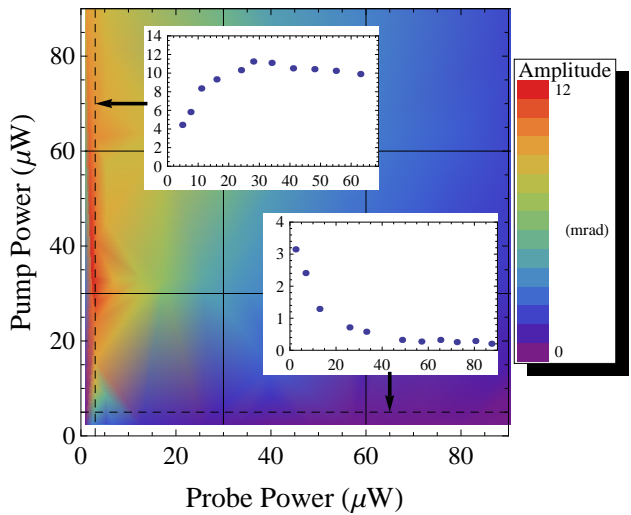}
    \caption{(Color online) AMOR signal as a function of the pump- and probe-light power. The data was measured at 850~mG. The top inset show a cross-section across the plot, i.e. amplitude pump-power dependence measured with a probe-intensity of 2~$\mu$W. Similarly, the bottom inset shows the data taken with varied probe power and fixed pump light intensity 4~$\mu$W.}
    \label{fig:Amplitude}
\end{figure}

As described above, the theoretical model developed in Sec.~\ref{sec:Theory} is valid only for low light powers; for higher light intensities such effects as ac Stark shift start to play an important role, e.g. by broadening of the AMOR resonances. Due to this fact not only the modulation frequency but also the pump and probe powers determine the amplitudes of the Zeeman coherences and hence the AMOR signals. In Fig.~\ref{fig:Amplitude} the amplitude of the AMOR signal is presented vs. the pump- and probe-light powers.
For low pump-light power the AMOR amplitude is small (see the upper inset to Fig.~\ref{fig:Amplitude}), which reflects low efficiency of the ground-state coherence generation. This efficiency, hence the AMOR-resonance amplitude, increases with the pump power and reaches its maximum at about 30~$\mu$W. Appearance of the maximum and further decrease of the amplitude results from the higher-order effects, such as saturation, hyperfine pumping, and repumping/regeneration of the existing Zeeman coherences. For instance, the hyperfine pumping leads to a decrease of a number of atoms in the $F=2$ ground state transferred to the $F=1$ ground state via spontaneous emission. Degradation of the AMOR signal is also associated with the applied modulation. Efficiency of the ground-state coherence generation and hence the number of atoms existing in a particular quantum state follows the light modulation. At low light intensities, it effects in a sinusoidal variation of numbers of atoms evolving with specific phases, which results in strong anisotropy of the medium. For more intense light, i.e., when saturation processes become significant, the efficiency does not reproduce the sinusoidal modulation pattern. In particular, higher harmonics of the modulation arise in the efficiency of coherence generation and number of atoms generated during successive pumping phases does not follow the sinusoidal dependence. It results in weaker anisotropy of the medium and a decrease of the AMOR-resonance amplitude for higher pump-light powers (Fig.~\ref{fig:Amplitude}).

The dependence of the AMOR signals on the probe-light intensity is different. As seen in the lower inset to Fig.~\ref{fig:Amplitude}, the amplitude of the AMOR resonance decreases with the probe-beam power on the whole accessible range of powers. This is caused by the probe light being resonant with the medium (the 80-MHz difference between the probe and pump frequencies caused by AOM is negligible relative to the Doppler broadening of the transition). In such a case, the probe perturbs the atoms; absorption of a photon from the probe beam results in a new quantum state which, in general, is different from the state created initially with the pump-light photon. In that way, the probe-light absorption decoheres the system and acts as additional relaxation mechanism reducing the AMOR-signal amplitude

\begin{figure}[htb!]
    \includegraphics[width=\columnwidth]{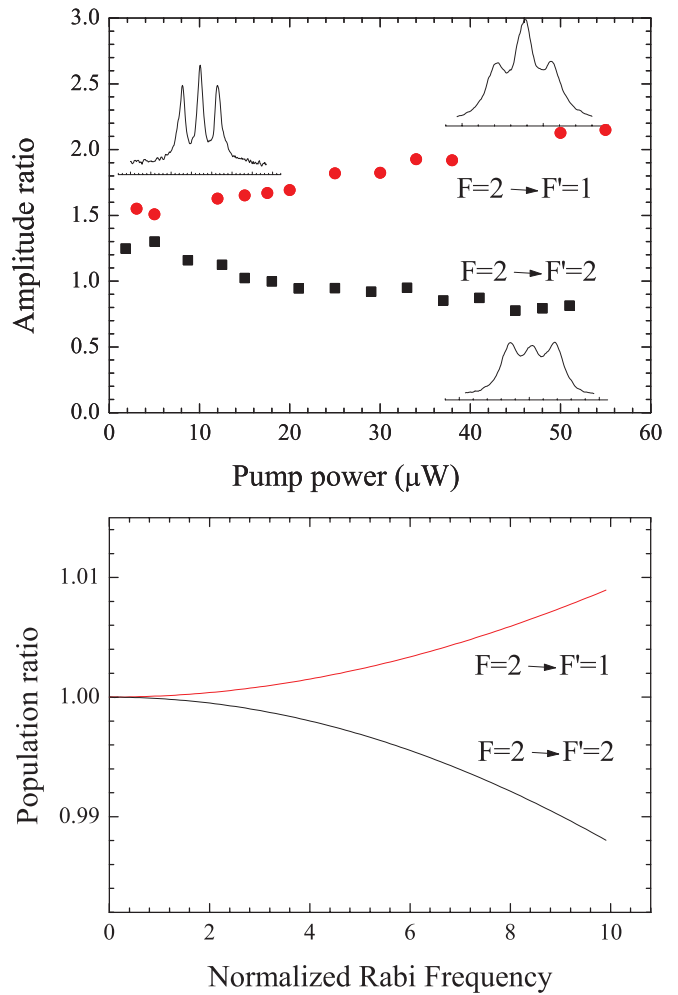}
    \caption{(Color online) (a) Ratio of the amplitude of the central of AMOR signal to the averaged amplitude of the side components of the signal vs. the light power, (b) ratio of the geometric mean of the $|-1\rangle$ and $|1\rangle$ sublevels population to the averaged geometric mean of the $|-2\rangle$ and $|0\rangle$ populations and $|0\rangle$ and $|2\rangle$ population vs. normalized Rabi frequency. Signals were measured at a magnetic field of 790~mG with the 1.2~$\mu$W~probe power while the calculations were performed for a single unmodulated light beam exploiting the rate equations. The red points and the red curve corresponds to the $F=2\rightarrow F'=1$ excitation while the black squares and black curve to the $F=2\rightarrow F'=2$ transition.}
    \label{fig:AmplitudeRatio}
\end{figure}

It was shown in Sec.~\ref{sec:Theory} that efficiency of the ground-state coherence generation strongly depends on the pump light tuning (dependence on $\Omega$). In particular, various pump-power dependences may be observed for the light coupling a given ground state with different excited states. Such an example is depicted in Fig.~\ref{fig:AmplitudeRatio}, which shows the ratio of the amplitude of the central resonance to the averaged amplitudes of the side resonances vs. the pump-light power for the $F=2\rightarrow F'=1$ and the $F=2\rightarrow F'=2$ transitions.
At low pump powers the AMOR signals observed at these two transitions are similar in shape and amplitudes, with well-resolved triple-component structures [see left-handed side inset to Fig.~\ref{fig:AmplitudeRatio}(a)]. However, the pump-power dependences of the resonances amplitudes measured at the two transitions are very distinct. While for the $F=2\rightarrow F'=1$ transition, the ratio increases with the pump-light power [see top right-handed inset to Fig.~\ref{fig:AmplitudeRatio}(a)], and the opposite dependence is observed at the other transition. In the latter case, the ratio decreases with the power [bottom right-handed inset to Fig.~\ref{fig:AmplitudeRatio}(a)] even below one when the side resonances have larger amplitudes than the central resonance. The observed dependences reflect the different behavior of the $\rho_{-11}$ coherence and the $\rho_{-20}$ and $\rho_{02}$ coherences at the two transitions. Among others, the behavior originates from optical pumping and population redistribution between Zeeman sublevels within a given hyperfine state. Since for the $F=2\rightarrow F'=1$ transition the maximal Clebsch-Gordan coefficients are those from the sublevels with maximal $m$ ($m=\pm F$), these states are most efficiently depopulated. Simultaneously, the depletion of the other sublevels is smaller, which effectively leads to an aligned state with the highest population in the $m=0$ sublevel and the lowest population in the $m=\pm 2$ sublevels. The change in the population distribution is reflected in the amplitudes of the Zeeman coherences associated with these states. For the $F=2\rightarrow F'=1$ transition and intense light, the lower amplitudes have the $\rho_{-20}$ and $\rho_{02}$ coherences, while the higher one has the $\rho_{-11}$ coherence. The opposite is true for the $F=2\rightarrow F'=2$ transition, while the $m=0$ state is most efficiently depleted.

In order to qualitatively verify the mechanism described above, the geometric means of  the population of the magnetic sublevels constituting the coherence were calculated vs. the pump intensity. Such a geometric mean of the two sublevels' populations sets an upper limit on the amplitude of the coherence between the sublevels (Schwarz inequality), $|\rho_{\alpha\beta}|\leq\sqrt{\rho_{\alpha\alpha}\rho_{\beta\beta}}$. Figure~\ref{fig:AmplitudeRatio}(b) shows the ratio of $\sqrt{\rho_{-1-1}\rho_{11}}$ to $(\sqrt{\rho_{-2-2}\rho_{00}}+\sqrt{\rho_{00}\rho_{22}})/2$ for the $F=2$ state coupled with light to the $F'=1,2$ states. Populations of the sublevels were calculated based on the rate equations using CW light and neglecting hyperfine optical pumping, saturation, etc. As seen in Fig.~\ref{fig:AmplitudeRatio}, the simulations qualitatively reproduce the observed dependence, i.e. increase of the ratio for the $F=2\rightarrow F'=1$ transition and decrease for the other transition.

\begin{figure}[h]
    \includegraphics[width=\columnwidth]{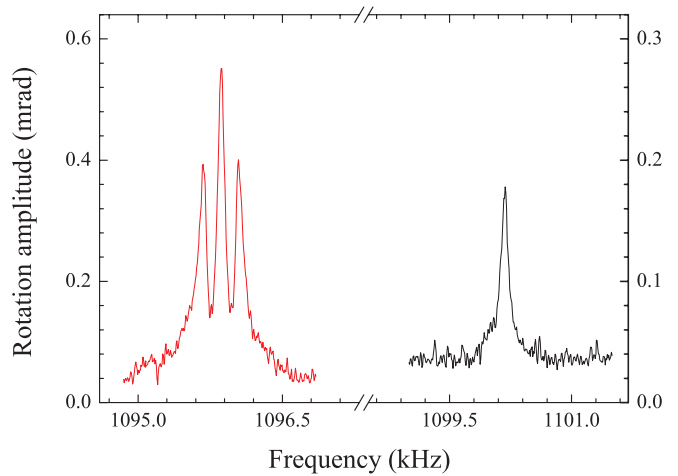}
    \caption{(Color online) AMOR signals recorded for light tuned to the $F=2\rightarrow F'=1$ transition (left curve) and the $F=1\rightarrow F'=1$ transition (right curve). The difference in the position of the AMOR resonances arises from nuclear contribution to the Land\'e factor. Signals were measured at a magnetic field of about $780$~mG and pump- and probe-light powers of 6~$\mu$W and 3~$\mu$W, respectively.}
    \label{fig:DifferentTransitions}
\end{figure}

In alkali atoms there are two hyperfine ground states supporting long-living quantum coherences. In particular, in $^{87}$Rb there are $F=1$ and $F=2$ ground states separated by about 6.8~GHz that can be selectively addressed by appropriate tuning of the pump and probe lasers. Figure \ref{fig:DifferentTransitions} presents the AMOR signals measured at the $F=2\rightarrow F'=1$ and $F=1\rightarrow F'=1$ transitions for the same set of experimental parameters.
As shown, the two signals are significantly different; not only different are the numbers of the AMOR resonances associated with the $\Delta m=2$ coherences (there is only one $\Delta m=2$ coherence in the $F=1$ ground state), but also their amplitudes are distinct. The latter difference originates from the transition probabilities and less efficient hyperfine pumping at the $F=1\rightarrow F'=1$ transition than at the $F=2\rightarrow F'=1$ one. Moreover, the positions of the AMOR resonances in a strong magnetic field are different. It results from the difference in the Land\'e factors for the two ground-state
\begin{equation}
    \begin{split}
        g_{F=2}=&-\frac{1}{4}g_J+\frac{5}{4}g_I,\\
        g_{F=1}=&\frac{1}{4}g_J+\frac{3}{4}g_I,
    \end{split}
    \label{eq:Splitting}
\end{equation}
where $g_J$ is the electron and $g_I$ the nuclear $g$-factor. Based on Eq.~(\ref{eq:Splitting}) it can be easily shown that although the splittings due to the electron spin are opposite and hence indistinguishable in the AMOR experiment, the nuclear-spin contributions are different, which leads to the separation of the resonance $\Delta\omega_m=2\times 2 g_I\mu_N B/\hbar$, where $\mu_N$ denotes the nuclear magneton . For a magnetic field of about 780~mG we predicted a splitting of about 4.35~kHz which is consistent with experimentally measured value of 4.32~kHz. This difference in frequencies of AMOR resonances for the transitions is important from a point of view of quantum-state engineering since it offers an additional way of controlling and modifying the quantum state by specific tuning of the frequency of an additional RF field \cite{Chalupczak2010Optical}.

\begin{figure}[htb!]
    \includegraphics[width=\columnwidth]{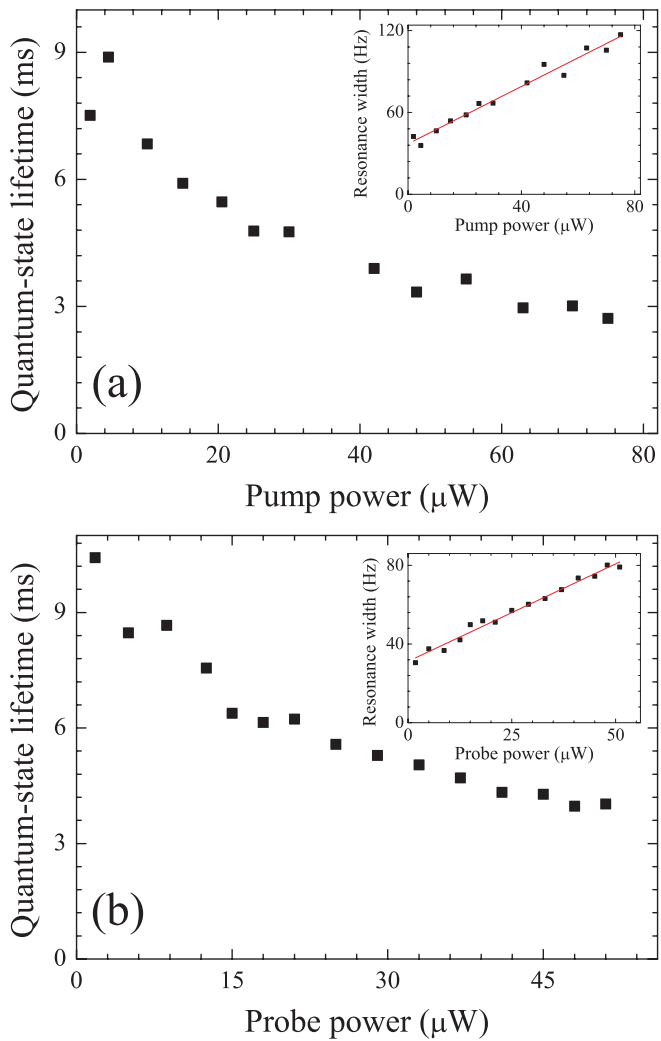}
    \caption{Lifetime of the Zeeman coherence between $|-1\rangle$ and $|1\rangle$ ground-state sublevels vs. pump- (a) and probe-light (b) powers. Increasing power of either of the light beams results in reduction of the lifetime of a quantum state which manifest as broadening of the AMOR resonances (see insets). The signals were measured for a magnetic field of about 780~$\mu$G with a probe power of 1.2~$\mu$W (a) and pump power of 2.2~$\mu$W. The laser was tuned to the center of the Doppler-broadened $F=2\rightarrow F'=1$ transition.}
    \label{fig:Lifetime}
\end{figure}

Dependences of the ground-state Zeeman-coherence lifetimes on the pump and probe powers are presented in Fig.~\ref{fig:Lifetime}.
The lifetimes $\tau$ were extracted from the AMOR resonance width as $\tau=1/\pi\delta\omega_m$, where $\delta\omega_m$ is the AMOR-resonance half-width at half maximum measured vs. the modulation frequency. Figure~\ref{fig:Lifetime} shows that raising the light power of either of the beams leads to broadening of the AMOR resonance (see insets) and shortening of the ground-state coherence lifetime. In order to calculate the coherence lifetime that is not affected by the light, we performed a series of measurements of the AMOR signals at different pump and probe powers and double-extrapolate the resonance width to zero light powers. The double-extrapolated lifetime of the coherences studied in this experiment is equal to 13.2(12)~ms, which is determined by three relaxation mechanisms: collisions with the uncoated surfaces, mainly in the cell stem containing a rubidium metal droplet, spin-exchange collisions between rubidium atoms \cite{Budker2005Antirelaxation}, and temperature-dependent dephasing collisions with the cell wall coating \cite{Pustelny2008Magentometry}. Using a simple mathematical model, the relaxation rate due to collisions with uncoated surfaces was estimated at the level of $\approx 2\pi\times 8.3$~s$^{-1}$. At the same time, the relaxation rate associated with spin-exchange collisions, that is calculated based on Ref.~\cite{Happer1972Review}, gives $2\pi\times 4.1$~s$^{-1}$. The other relaxation channel most likely are dephasing collisions of atoms with the coating.

\section{Summary and conclusions\label{sec:Conclusions}}

We have analyzed the possibility of generating the quantum superpositions of ground-state Zeeman sublevels differing in the magnetic quantum numbers by 2. Since in the $F>1$ state the sublevels split nonlinearly with a magnetic field (nonlinear Zeeman effect), selective generation of coherences between specific sublevels by appropriate tuning of the modulation frequency is possible. In particular, it was shown that for the magnetic fields such that nonlinear magnetic-sublevel splitting exceeds the ground-state relaxation rate, selective addressing of the $\rho_{-20}$, $\rho_{-11}$, and $\rho_{02}$ coherences is possible. Efficiency of the coherence generation versus different experimental parameters, such as modulation frequency, pump- and probe-light power, light frequency was analyzed. We have shown that in our experimental setup the lifetime of the coherences exceeds 10~ms. Such a long coherence lifetime opens interesting possibilities for application of the coherences in quantum-state engineering and quantum computation. In this context particularly interesting seems to be the ability of modification of atomic quantum state by application of external field, e.g. magnetic and/or RF field.

\begin{acknowledgments}
The authors would like to express their gratitude to Andrew Park and Simon Rochester for stimulating discussions. The work was supported by the Polish Ministry of Science and Higher Education (grants N N202 074135 and N N202 175935). Part of the work was operated within the Foundation for Polish Science Team Programme co-financed by the EU.
\end{acknowledgments}


\begin{thebibliography}{99}
\bibitem{FootnoteNeglectingDecay} Since out experiment was performed with low pump and probe powers ($\Omega^2_{pu,pr}\ll \Gamma\gamma$), the theoretical considerations neglect processes associated with spontaneous decay of the excited state, i.e. population and Zeeman coherence transfer from the excited state to the ground state. We have also neglected ground-state relaxation due to such processes as spin-exchange collisions. Including these preocesses in our consideration would require introduction additional terms in Eq.~(\ref{eq:LiouvilleEquation}).
\bibitem{FootnoteRWA} Note that the formulae are derived within the rotating-wave approximation.
\bibitem{FootnoteModulation} It should be noted that in case of other modulation (different modulation waveforms or modulation depths) Eq.~(\ref{eq:IntensityModulation}) need to be expanded in a Fourier series with respect to $\omega_m$ and then ought to be considered analogically.
\bibitem{FootnoteMeasuringSplitting} The splitting of the AMOR resonance is measured by fitting three Lorentz curves with independent amplitudes and positions to the amplitude of the AMOR signal (see, for example, Fig.~\ref{fig:AMORsignal}). At low magnetic fields, when the components are barely resolved, the fitting was performed by presetting amplitude ratio of all AMOR resonances.
\end{thebibliography}
\end{document}